\documentclass[letterpaper,twocolumn,11pt]{article}

\newlength\titlebox
\usepackage{times}
\usepackage{helvet}
\usepackage{courier}
\usepackage{url}
\usepackage{graphicx}
\usepackage{booktabs}
\usepackage{multirow,rotating}
\usepackage[usenames,dvipsnames]{color}
\usepackage{amsmath}

\usepackage[]{authblk}  

\usepackage[font=small]{caption}
\usepackage{abstract}
\date{April 4, 2011}
\usepackage[compact]{titlesec}

\def \Vb  {V_\mathrm{B}}
\def \Vn  {V_\mathrm{N}}
\def \Vl  {V_\mathrm{L}}
\def \Ebl {E_\mathrm{BL}}
\def \Enl {E_\mathrm{NL}}
\def \var {\mathrm{Var}}
\def \Dem {\mathrm{D}}
\def \Rep {\mathrm{R}}
\def \pD {p_\mathrm{D}}
\def \pR {p_\mathrm{R}}

\def \Tpop {$\Theta_{\mathrm{pop}}$}
\def \Tcon {$\Theta_{\mathrm{con}}$}
\def \nn {News}
\def \bb {Blogs}

\newcommand{\vrtLbl}[2]{\multirow{#2}{1ex}{\begin{sideways}#1\end{sideways}}} 
\newcommand{\shortcite}[1]{\cite{#1}}

\title{More Voices Than Ever? Quantifying Media Bias in Networks}

\author[1,4,*]{Yu-Ru Lin}
\author[3,5]{James P.~Bagrow}
\author[1,2,4]{David Lazer}
\affil[1]{College of Computer and Information Science, Northeastern University, Boston, MA 02115, USA}
\affil[2]{Department of Political Science, Northeastern University, Boston, MA 02115, USA}  
\affil[3]{Department of Physics, Northeastern University, Boston, MA 02115, USA}  
\affil[4]{Institute for Quantitative Social Science, Harvard University, Cambridge, MA 02138, USA}  
\affil[5]{Dana-Farber Cancer Institute, Harvard University, Boston, MA 02215, USA}  
\affil[*]{To whom correspondence should be addressed. Email: \protect\url{yuruliny@gmail.com}}

\begin{document} 

\twocolumn[
    \maketitle
    \vspace{-3em}
    \begin{onecolabstract}
        \vspace{-1em}
        Social media, such as blogs, are often seen as democratic entities that allow
        more voices to be heard than the conventional mass or elite media. Some also
        feel that social media exhibits a balancing force against the arguably slanted
        elite media. A systematic comparison between social and mainstream media is
        necessary but challenging due to the scale and dynamic nature of modern
        communication. Here we propose empirical measures to quantify the extent and
        dynamics of social (blog) and mainstream (news) media bias. We focus on a
        particular form of bias---coverage quantity---as applied to stories about the
        111th US Congress. We compare observed coverage of Members of Congress against a
        null model of unbiased coverage, testing for biases with respect to political
        party, popular front runners, regions of the country, and more. Our measures
        suggest distinct characteristics in news and blog media. A simple generative
        model, in agreement with data, reveals differences in the process of coverage
        selection between the two media.
    \end{onecolabstract}
]
  
\noindent ``In the end, we'll have more voices and more options." 
\vspace{-1.5em}\begin{flushright}-- Dan Gillmor, \textit{We the media}\end{flushright}

\section{Introduction} 
Gillmor \shortcite{gillmor2004we} envisioned social media, powered by the growth
of the Internet and related technologies, as a form of grassroots journalism
that blurs the line between producers and consumers and changes how information
and opinions are distributed. He argued that ``the communication network itself
will be a medium for everyone's voice, not just the few who can afford to buy
multimillion-dollar printing presses, launch satellites, or win the government's
permission to squat on the public airways." This view has been embraced by
activists who consider social media as a balancing force to the conventionally
assumed slanted or biased elite media.  Indeed, social media can be used by
underprivileged citizens, promising a profound impact and a healthy democracy.

Many believe that the mainstream media is slanted, but disagree about the
\emph{direction} of slant.  The conventional belief about media bias has held
for decades, but attempts at developing objective measurement have only recently
begun. The study by Groseclose and Milyo \shortcite{groseclose2005measure}
showed the presence of bias in mass media (cable and print news) and new media
(Internet websites, etc.). Their results, despite receiving criticism, are
fairly consistent with conventional wisdom. On the other hand, researchers have
observed an ``echo chamber'' effect within the new media -- people select
particular news to reinforce their existing beliefs and attitudes. Iyengar and
Hahn \shortcite{iyengar2009red} argued that such selective exposure is
especially likely in the new media environment due to information overload. With
search, filtering, and communication technologies, people can easily discover
and disseminate information that are supportive or consistent with their
existing beliefs.

Do social media exhibit more or less bias than mass media and, if so, to what
extent? Identifying media bias is challenging for a number of reasons. First,
bias is not easy to observe. It has been recognized that ``bias is in the eyes
of the beholder" meaning that, e.g., conservatives tend to believe that there is
a liberal bias in the media while liberals tend to believe there is a
conservative bias \cite{groseclose2005measure,yano2010shedding}. Hence, finding
textual indicators of bias is difficult, if not impossible. Second, the
assessment of bias usually implies knowing what ``fairness" would be, which may
not be available or consistent across different viewpoints. Third,
Internet-based communication promises easy, inexpensive, and instant information
distribution, which not only increases the number of online media outlets, but
also the amount and frequency of information and opinions delivered through
these outlets.  The scale and dynamic nature of today's communication should be
accounted for.

In this paper, our major contribution is that we propose empirical measures to
quantify the extent and dynamics of ``bias'' in mainstream and social media
(hereafter referred to as \textit{\nn} and \textit{\bb}, respectively). Our
measurements are not normative judgment, but examine bias by looking at the
attributes of those being mentioned, against a null model of ``unbiased''
coverage.
We focus on the number of times a member of the 111th US congress was
\textit{referenced},  and study the distribution and dynamics of the references
within a large set of media outlets. We consider ``the unbiased" as a
configurable baseline distribution and measure how the observed coverage
deviates from this baseline, with the measurement uncertainty of observations
taken into account. We demonstrate bias measures for slants in favor of specific
political parties, popular front-runners, or certain geographical regions. 
Using these measures to examine newly collected data, we have observed distinct
characteristics of how \nn{} and \bb{} cover the US congress. Our analysis of
party and ideological bias indicates that \bb{} are not significantly less
slanted than \nn{}. However, their slant orientations are more sensitive to
exogenous factors such as national elections. In addition, blogs' interests are
less concentrated on particular front-runners or regions than news outlets.

While our measures are independent of content, we further investigate two
aspects of the content related to our measures: the hyperlinks embedded in
articles and sentiments detected from the articles. The hyperlink patterns
suggest that outlets with a Democrat-slant (D-slant for short) are more likely
to cite each other than outlets with a Republican-slant (R-slant). The sentiment
analysis suggests there is a weak correlation between negative sentiments and
our measures.

To better understand the distinctive slant structures between the two media, we
propose to use a simple ``wealth allotment'' model to explain how legislators
gain attention (references) from different media. The results about blog media's
inclination to a rich-get-richer mechanism indicates they are more likely to
echo what others have mentioned. 
This observation does not contradict our measures of bias -- compared with news
media, blogs are weaker adherents to particular parties, front-runners or
regions but are more susceptible to the network and exogenous factors.

The rest of this paper is organized as follows. We first discuss related work,
followed by the details of our collected data.  We then detail the different
types of coverage bias and how to quantify them and then examine the results,
both structurally (via hyperlinking) and textually (via text-based sentiment
analysis).  Finally, we present a simple generative model of media coverage and
conclude with a discussion of open issues and future work.

\section{Related Work}
Concerns about mainstream media bias have been a controversial and critical
subject in journalism due to the media's power to shape a democratic society.
Studies on media bias can involve surveys and interviews
\cite{lichter1986media}, and content analysis \cite{eldridge1995glasgow}, as
well as theoretical models such as structural economic causes. Apart from these
qualitative arguments, Groseclose and Milyo \shortcite{groseclose2005measure}
proposed a media bias measure that counts how often a particular media outlet
cites various think tanks and policy groups.

There have been controversial responses to prior studies, and the origin in part
lies in the difficulty to separate the recognition of bias from the belief of
bias. A dependence on viewers' beliefs has been observed in studies
\cite{groseclose2005measure,yano2010shedding}, which is relevant to the theories
on how supply-side forces or profit-related factors cause slants in media
\cite{mullainathan2005market,gentzkow2010drives}. Because of such a dependency,
computationally identifying bias from media content remains an emerging research
topic, and requires insights from other language analysis studies such as
sentiment analysis \cite{pang2008opinion} or partisan features in texts
\cite{monroe2008fightin,gentzkow2010drives}.

While mass media have the ability to affect the public's interests, social media
represent large samples of expression from both influencers and those being
influenced. Hence the ``crowd voice'' collected in social media has attracted
considerable research. The viral behavior and predictive power of social media
in response to politics, the economy and other areas has been examined in recent
studies \cite{leskovec2009meme,o2010tweets}. For example,
Leskovec et al.~\shortcite{leskovec2009meme} tracked the traversal of ``memes''
based on short distinctive phrases echoed by online news and blogs over time.
Another work by
O'Connor et al.~\shortcite{o2010tweets} studied the relationship between tweet
sentiments and polls in order to examine how the sentiments expressesed in the
Twitter microblogging social media can be used as political or economic
indicators.

In this paper, we do not attempt to tackle the computationally difficult task of
identifying bias in media text. Instead, we study the characteristics of the two
media based on purely quantitative measures independent of media content. We are
interested in studying the role of today's social media, and we hope our
analysis will contribute to the growing understanding of this subject.

\section{Data Model}
\subsubsection{Data Collection}
Our data is based on RSS feeds aggregated by
OpenCongress\footnote{\url{www.opencongress.org}}\footnote{OpenCongress uses
    Daylife (\url{www.daylife.com}) and Technorati (\url{technorati.com}) to
    aggregate articles from these feeds. The possible selection biases in these
    filtering processes are not considered in this paper.}. OpenCongress is a
non-profit, non-partisan public resource website that brings together official
government data with timely information about what is happening in Congress. We
continuously monitor and collect the OpenCongress RSS feeds for each individual
member of Congress\footnote{An example news/blog coverage feed can be found at
    \url{http://www.opencongress.org/people/news_blogs/300075_Lisa_Murkowski}}.
This paper examines \nn{} and \bb{} coverage about the 111th US Congress, both
Senators and Representatives. The dataset spans from September 1 to January 4,
covering the 2010 mid-term election on November 2.

\begin{figure}\centering
    {\includegraphics[trim=0 10 0 5]{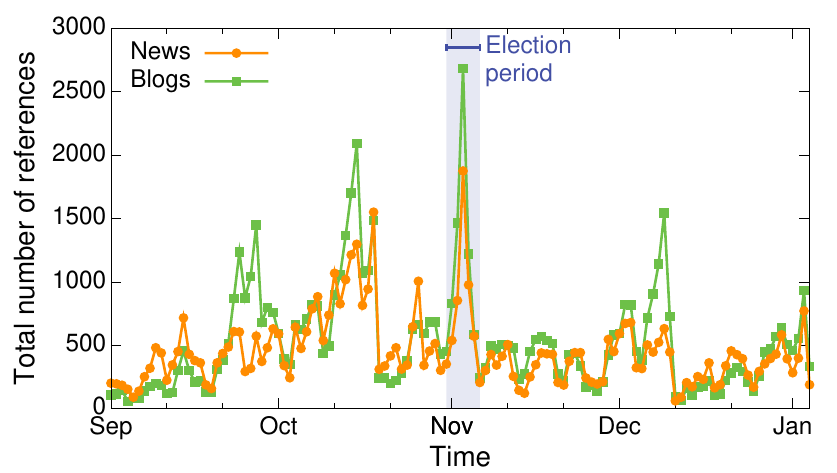}}
    \caption{\label{volume_w}The volume (total number of news articles or blog
        posts) over time. The highest peak corresponds to the mid-term
        election.}
\end{figure}

Figure~\ref{volume_w} shows the volume (total number of news articles or blog
posts) over time in this dataset. The central peak corresponds to the mid-term
election. In total, there are 57,221 news articles and 66,830 blog posts being
collected in the four-month period.

\subsubsection{Networked Data Model}
\begin{figure}[t!]\centering
	{\includegraphics[width=.7\columnwidth]{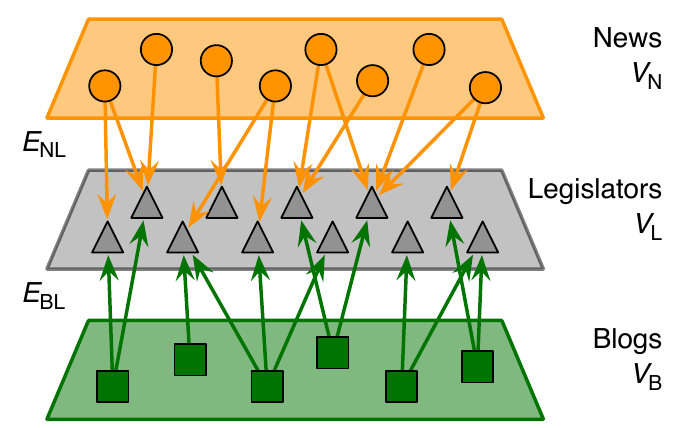}}
    \caption{\label{fig_tripartite}The networked data model. There are three
        types of nodes: news outlets, blog outlets and legislators. An edge
        pointing toward a legislator represents each time an outlet references
        that legislator in an article or post.}
\end{figure}

We study the structure of the two media by constructing a modal network
containing different types of nodes and edges. The network structure is
illustrated in Fig.~\ref{fig_tripartite}. More specifically, we have:
\begin{description}
    \item[Nodes] There are three sets of nodes: a news set, denoted by $\Vn$,
        that contains 5,149 news outlets, a blog set $\Vb$ of 19,693
        blogs\footnote{We also have a small number of blogs hosted by mass media
            news outlets, e.g. CNN (blog). This paper does not include analysis
            of such blogs.}, and a legislator set $\Vl$ that covers 530
        lawmakers.
    \item[Edges] Each edge $e_{ik}$ records when media outlet $i$ publishes an
        article referencing legislator $k$. We extract 64,222 such edges in
        46,501 news articles, denoted as edge set $\Enl$, and 91,837  edges in
        62,301 blog posts, denoted as $\Ebl$. Edges are associated with
        timestamps and texts.
    \item[Node attributes] For legislators, we record attributes such as party,
        district, etc., based on the legislators' profiles and external data
        sources.
\end{description}
While we focus on ``reference" or citation edges, this networked model can also
include other types of edges, e.g.~hyperlinks between outlets, voting
preferences among legislators, etc.

\section{Types of Bias}
In journalism, the term ``media bias" refers to the selection of which events
and stories are reported and how they are covered within the mass media. The
most commonly discussed biases include reporting that supports (or attacks)
particular political parties, candidates, ideologies, corporations, races, etc.
In this paper, we begin with perhaps the simplest form of measurable bias -- the
distribution of coverage quantity, i.e. how many times an entity of interest is
referenced by a media outlet. We argue that, regardless of a positive or
negative stance towards an entity, an imbalanced \emph{quantity} of coverage, if
present, is itself a form of bias\footnote{Our view on the meaningfulness of a
    measurement based solely on quantity is similar to the study of Groseclose
    and Milyo \shortcite{groseclose2005measure}.}.

An outlet's references can be biased in a number of ways:
\begin{description}
    \item[Party] References are focused on a particular political party.
    \item[Front-runner] References are concentrated on a few legislators who we
        term ``front-runners", while the majority of legislators receive little
        or no attention.
    \item[Region] References focus on certain geographical locations.
    \item[Ideology] An ideology is a collection of ideas spanning the political
        spectrum. Ideological bias indicates that frequently referenced
        legislators favor certain ideological tendencies.  \item[Gender] The
        preference towards covering legislators of one gender.
\end{description}
We discuss how to measure different types of bias in a unified model. Other
types of bias, such as those in favor of a particular race or ethnic group, can
also be measured through our method.

Based on the measurements associated with individual media outlets, we derive
system-wide bias measures that allow us to characterize and compare the bias
structure between the news and blog media.

\section{Quantifying Bias}
In this section, we describe our method for quantifying and comparing bias in
\nn{} and \bb{}. 

\subsubsection{Notation} Let $n_{ik}^c$ be number of times media outlet $i$
references legislators in group $k$, where $c\in\{\mbox{\nn, \bb}\}$ is the
media category ($c$ is omitted when there is no need to distinguish the
categories). In the case of measuring party bias, $k\in\{\Dem, \Rep\}$ indicates
the Democratic or Republican political parties. Let $n_i=\sum_k{n_{ik}}$ be the
total number of references made by outlet $i$. We begin with a specific case --
measuring the two-party bias, and then describe a more general model for
measuring other types of bias.

\subsection{Party Slant}
A naive approach for measuring an outlet's biased coverage of two political
parties is to compare the number of times members in each party are referenced.
The ratio of the reference counts of one party against the other may be used to
compare outlets that reference different parties with different frequencies.
There are two issues with this approach: (i) this ratio may lack  statistical
significance for some outlets, and (ii) it assumes that fair coverage of the two
parties requires roughly equal quantities of references to each.

To resolve these issues, we use the \textit{log-odds-ratio} as follows. We
define $\theta_{ik}$, the ``slant score'' of outlet $i$ to party $k$, as
\begin{equation}\label{eqn:logodds}
	\theta_{ik} = \log(\mbox{odds-ratio})=\log\left(\frac{n_{ik}/(n_i-n_{ik})}{p_k/(1-p_k)}\right),
\end{equation}
where $p_k$ is the \textit{baseline probability} that $i$ refers to $k$, and
here we assume this variable is fixed for all $i$. The advantage of having such
a baseline probability is that ``fairness" become configurable. For example, one
can consider fairness as a 50-50 chance to reference either party (i.e.
$\pD=\pR=0.5$). One can also define $\pD=0.6$ since roughly 60\% of the studied
legislators are Democrats. No matter what baseline probability is given, we have
a simple interpretation: $\theta = 0$ means no bias w.r.t that baseline. In this
two-party case, we take $\theta_i\equiv\theta_{ik}$, with $k=\mathrm{D}$, and
$\theta_{i}>0$ means outlet $i$ is more likely to be D-slanted. A slant score
with value $\alpha$ can be interpreted as follows: the number of times outlet
$i$ references Democratic legislators is $e^\alpha$ times more than if those
references followed the baseline.

The slant score's variance is given by the Mantel-Haenszel estimator
\shortcite{mantel1959statistical}:
\begin{equation}
	\var(\theta_i)=\frac1{n_{ik}}+\frac1{n_i-n_{ik}}+\frac1{n_ip_k}+\frac1{n_i(1-p_k)}.
\end{equation}
The variance gives the significance of the slant score measure, which relies on
the number of observations ($n_i$ and $n_{ik}$) we have for each outlet.

\begin{figure}
	\centering
	\includegraphics[trim=0 5 0 0,width=1\columnwidth]{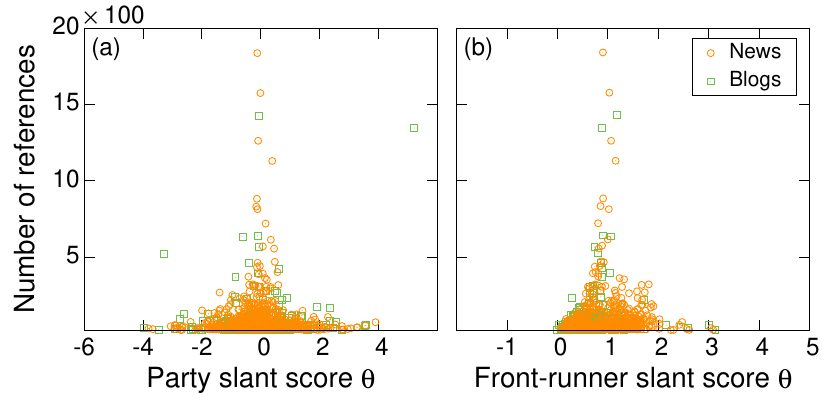}
    \caption{\label{party-scatter}The scatter plot of number of references
        (observations) against party (left) and front-runner (right) slant
        scores for \nn{} and \bb{}. Outlets with less than 20 articles are not
        shown.}
\end{figure}

Figure \ref{party-scatter} (a) shows the number of references as a function of
party slant scores for outlets with more than 20 articles in our dataset. The
distribution of outlets' slant scores appears to be roughly symmetric in both
directions, and outlets making more references tend to be less slanted. Table
\ref{top-outlets} lists the slant scores for some major news outlets and the
most slanted blogs.

\begin{table}
    \centering 
    \caption{\label{top-outlets}Slant scores $\theta$ for major news outlets and
        most slanted blogs. For party slant, a positive (negative) score means
        the outlet is likely to be D-slanted (R-slanted). For front-runner and
        regional slant, a larger score indicates the outlet is more focused on
        few particular legislators or states.} \tiny
    \begin{tabular}{@{}crrr@{}}
    \toprule              
                          & Party ($\theta$)                   & Front-runner ($\theta$)          & Region   ($\theta$)                \\
    \midrule                                            
        \vrtLbl{News}{16} & nbc                  (0.51) & washington post         (1.03) & los angeles times            (1.30) \\
                          & new york times       (0.07) & cnn                     (1.02) & nbc                          (1.19) \\ 
                          & washington post     (-0.01) & fox                     (0.91) & cbs                          (1.12) \\
                          & abc                 (-0.03) & wall street journal     (0.86) & cnn                          (1.04) \\
                          & cbs                 (-0.03) & cbs                     (0.84) & washington times             (1.00) \\
                          & los angeles times   (-0.07) & nbc                     (0.83) & u.s. news                    (0.98) \\
                          & newshour            (-0.10) & los angeles times       (0.82) & wall street journal          (0.96) \\
                          & cnn                 (-0.11) & msnbc                   (0.74) & usa today                    (0.96) \\
                          & fox                 (-0.13) & u.s. news               (0.71) & washington post              (0.95) \\
                          & npr                 (-0.14) & new york times          (0.70) & msnbc                        (0.92) \\
                          & wall street journal (-0.15) & washington times        (0.70) & npr                          (0.92) \\
                          & u.s. news           (-0.22) & usa today               (0.66) & new york times               (0.89) \\
                          & bbc                 (-0.38) & npr                     (0.64) & abc                          (0.87) \\
                          & usa today           (-0.39) & abc                     (0.61) & fox                          (0.84) \\
                          & msnbc               (-0.39) & newshour                (0.32) & newshour                     (0.78) \\
                          & washington times    (-0.96) & bbc                     (0.00) & bbc                          (0.20) \\
    \midrule                                                                                  
        \vrtLbl{Blogs}{5} & dissenting times     (5.22) & arlnow.com              (9.41) & blue jersey                  (8.32) \\ 
                          & cool wicked stuff    (3.89) & janesville              (9.05) & [...] virginia politics      (7.86) \\ 
                          & justicedenied13501   (3.58) & take back idaho's [...] (8.84) & politics on the hudson       (7.34) \\ 
                          & polifrog.com         (3.54) & moral science club      (8.84) & calwatchdog                  (7.23) \\ 
                          & dennis miller        (3.46) & murray for congress     (8.67) & staradvertiser [...]         (7.19) \\ 
    \bottomrule
    \end{tabular}
\end{table}

\subsubsection{Summary statistics} 
In order to characterize the overall bias within a media, we derive a
system-wide bias measure based on the individual outlets' measures. We use a
\textit{random effect} model, which assumes not only variation within each
outlet, but also variation across different outlets in the system. More
specifically, the model assumes that the slant scores for $n$ outlets
$(\theta_1, \ldots,\theta_n)$ are sampled from $\mathcal{N}(\theta,\tau^2)$, and
there are two sources of variation: the variance between outlets $\tau^2$ and
the variance within outlets $\sigma^2$. Hence, the model is given by
\begin{equation}
	\hat\theta_i\sim \mathcal{N}(\theta,\sigma^2+\tau^2).
\end{equation}
We use the DerSimonian-Laird estimator \shortcite{dersimonian1986meta} to obtain
$\theta^*$ and $\var(\theta^*)$, where $\theta^*$ is the asymptotically unbiased
estimator for $\theta$. The media-wide \textit{collective party slant score},
$\Theta$, is defined as $\Theta\equiv\theta^*$ with a
$\pm1.96\sqrt{\var(\theta^*)}$ confidence interval.

Table \ref{slant-scores} summarizes slants with respect to different baselines.
The measure \Tcon{} is based on the party composition of members in Congress,
and \Tpop{} is based on the fraction of the US population represented by the
legislators (in each party). The statistical significance of each measure is
represented by the variance. Note that in this two-party case, a different
baseline can be obtained simply by shifting the score. For example, if one
chooses to use $\pD=\pR=0.5$ as the baseline probability, the measure
$\Theta_{0.5}$ can be calculated from \Tcon{} by adding
$\log(\frac{\pD}{1-\pD})\approx 0.405$ (where in terms of Congress composition
$\pD\approx0.6$). 

\begin{table}[t]
	\centering 
    \caption{\label{slant-scores}The collective slant scores. Parenthetical
        values indicate standard deviation of the measured slant
        score. } \tiny
	\begin{tabular}{r r cc cc }
	\toprule
	       &    &  \multicolumn{2}{c}{House} & \multicolumn{2}{c}{Senate} \\
	\cmidrule(r){3-4}  \cmidrule(r){5-6} 
	       &    & \Tcon & \Tpop & \Tcon & \Tpop  \\
	\midrule
	\multirow{2}{*}{Party}        & News   & -0.02 (0.02) & -0.06 (0.02)   & -0.22 (0.03) & -0.45 (0.04)    \\
                                  & Blogs  & -0.11 (0.02) & -0.15 (0.02)   & -0.18 (0.04) & -0.41 (0.04)    \\
	\midrule                                                                                              
	\multirow{2}{*}{Ideology}     & News   & -0.05 (0.02) & -0.08 (0.02)   & -0.19 (0.04) & -0.45 (0.04)    \\
                                  & Blogs  & -0.16 (0.02) & -0.19 (0.02)   & -0.12 (0.04) & -0.39 (0.04)    \\
	\midrule                                                                                              
	\multirow{2}{*}{Gender}       & News   & -0.26 (0.04) &  0.07 (0.03)   & -0.28 (0.06) &  0.45 (0.05)    \\
                                  & Blogs  & -0.29 (0.04) &  0.03 (0.04)   & -0.32 (0.07) &  0.41 (0.06)    \\
	\midrule                                                                                              
	Front-                        & News   &  0.68 (0.01) &  0.60 (0.01)   &  0.66 (0.02) &  0.55 (0.03)    \\
    runner                        & Blogs  &  0.33 (0.01) &  0.23 (0.01)   &  0.39 (0.02) &  0.29 (0.03)    \\
	\midrule                                                                                              
	\multirow{2}{*}{Region}       & News   &  0.97 (0.01) & -0.13 (0.01)   &  0.76 (0.01) &  0.45 (0.03)    \\
                                  & Blogs  &  0.61 (0.01) & -0.21 (0.02)   &  0.44 (0.02) &  0.18 (0.03)    \\
	\bottomrule
	\end{tabular}
\end{table}

We also separate our measures for referencing members of the House and Senate to
see if outlets exhibit different slants when covering the two chambers.
Evaluated on the party percentage baseline, both media show R-slant when
referencing Senators, and blogs are more R-slanted when referencing members of
the House. Hence \bb{} are overall more R-slanted than \nn{}. This
interpretation depends on what baseline is chosen, however. For example, if we
choose to use the 50-50 convention, both media become D-slanted. However, it is
important to note that the absolute difference between the bias measures for the
two media do not change with baseline.

\subsection{Slant Dynamics}
\begin{figure}
  \centering
  {\includegraphics[trim=0 15 0 24]{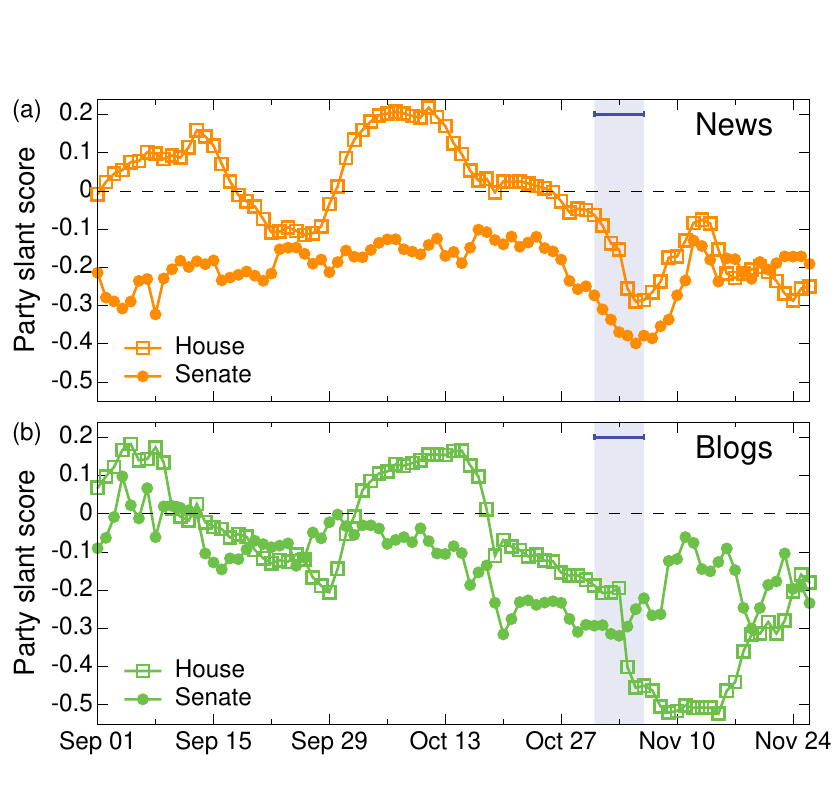}}
  \caption{\label{party-slant-dynamics}Slant score as a function of time.
      Overall, the media, especially \bb{}, become more R-slanted after the 2010
      election.}
\end{figure}

To study how media bias may change over time, we calculate the slant scores
using references made during running windows. We measure $\Theta(t,w)$ as a
function of time $t$ and window length $w$. Figure \ref{party-slant-dynamics}
shows the temporal slant scores for the two media during the four-month period,
based on a $w=\mbox{2-week}$ running window. The slant of both media changes
slightly after the mid-term election: Compared with their pre-election slants,
\nn{} become slightly more R-slanted when referencing Senators and \bb{} are
more R-slanted when referencing Representatives. Overall, the media, especially
\bb{}, become more R-slanted after election.  This is reasonable due to the
Republican victories.

\begin{figure}[t!]
	\centering
    {\includegraphics[trim=0 6 0 3.6]{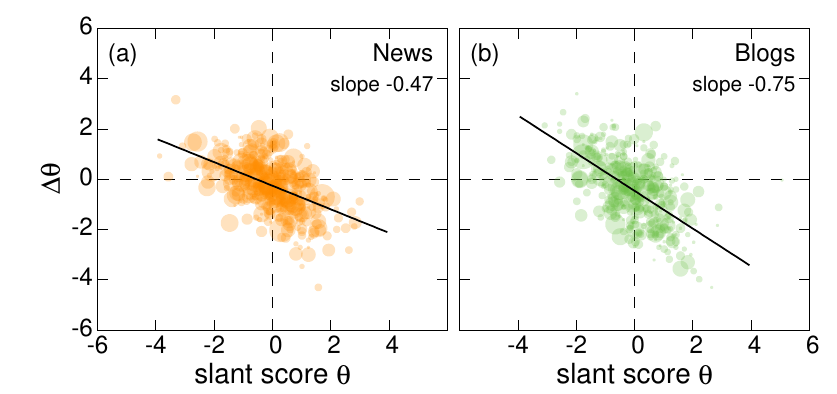}}
    \caption{\label{party-slant-diff}Media outlets are slightly shifting towards
        the other side after election. The majority of news outlets become
        slightly more R-slanted. For blogs, originally D-slanted blogs become
        more R-slanted.  Each point represents a media outlet.}
\end{figure}

These results raise an important question: do the majority of outlets become
more R-slanted after the election, or do R-slanted outlets become more active
while D-slanted outlets become quieter? To examine what caused the slant change
we plot in Fig.~\ref{party-slant-diff} the change in slant score $\Delta
\theta_i = \theta_i(t_2) - \theta_i(t_1)$, where $t_1\in\mbox{[Sep.~1,
    Oct.~30]}$ and $t_2\in\mbox{[Nov.~7, Jan.~4]}$, for each outlet against its
slant score before the election. (Point size indicates the amount of references
observed after the election.) We use a linear regression to quantify the slant
change. Surprisingly, we see media outlets shifted slightly toward the other
side after the election regardless of their original slants, but overall the
originally D-slanted outlets become more R-slanted.

\subsection{Front-Runner Slant}
To evaluate whether or not the media pay excessive attention on popular
front-runners, we extend the dichotomous-outcome measure used in the previous
section.  We consider a generalization of the odds ratio proposed by Agresti
\shortcite{agresti1980generalized}.

Let $n_{ik}^c$ now be the number of times outlet $i$ refers to the $k$-th
legislator, where $c\in\{\mbox{\nn, \bb}\}$ as before, and $k\in\{1,2,...,L\}$
is the \textit{rank index} for one of the $L$ legislators, ordered by the number
of references received from outlet $i$. We can replace $n_{ik}$ by the sample
proportion $p_{ik}=n_{ik} / n_i$.  The slant score $\theta_i$ of outlet $i$ is
defined by a generalized log-odds-ratio: 
\begin{equation}\label{glor}
	\theta_i = \log\left(\frac{\sum_{j>k}{p_{ik}p_j}}{\sum_{j<k}{p_{ik}p_j}}\right),
\end{equation}
where $p_j$ is, again, the baseline probability that $i$ refers to the $j$-th
legislator, and the $\{p_j\}$ can be chosen to be uniform or any other
distribution. For convenience we commonly fix the baseline distribution for all
$i$. 

When $L=2$, Eq.~\ref{glor} reduces to a dichotomous-outcome log-odds-ratio
measure similar to Eq.~\ref{eqn:logodds}. When $L>2$ and the $\{p_j\}$ are not
uniform, changing to a different baseline is not a simple linear shift.  With
Eq.~\ref{glor}, a slant score with value $\alpha$ can be interpreted as follows:
the number of times outlet $i$ mentions high ranked legislators is $e^{\alpha}$
times more than if the legislators were ranked according to their baseline
probabilities.

The variance in the slant score is now given by \cite{agresti1980generalized}:
\begin{equation}\label{glor-var}
	\var(\theta_i)= \frac{\sum_j p_{ij}\left(\alpha_{ij} \right)^2 + \sum_j p_j\left(\beta_{ij}\right)^2}
{n_i\left(\sum_{k>j}p_{ik}p_j\right)^2}
\end{equation}
where
\begin{equation*}
	\alpha_{ij}=\theta_i \sum_{k<j}p_k - \sum_{k>j}p_k, ~~~~
	\beta_{ij}=\theta_i\sum_{k>j}p_{ik} - \sum_{k<j}p_{ik}.
\end{equation*}

Figure \ref{party-scatter} (b) plots the number of references (observations)
against front-runner slant scores for media and blog outlets with more than 20
posts in our dataset. We expect the frontrunner slant scores to be mostly
positive since the legislators are already ranked by popularity ($n_{ik}$).

The system-wide frontrunner slant score for both news and blog media can be
calculated as before. Table \ref{slant-scores} summarizes front-runner slants
with respect to various baselines. Note that the two media show different biases
when referencing the two chambers: Blogs are more front-slanted than news about
Senators, while news outlets are more front-slanted when referencing
Representatives.

\subsection{Other Types of Slant}
\subsubsection{Ideology}
The concept of ideology is closely related to that of political party -- members
of the same party usually share similar or less contradictory ideologies.  We
study the ideological bias using a method similar to the party slant analysis.
We first locate each legislator relative to an identifiable ideological
orientation such as left or right, 
and then use the dichotomous-outcome measure to obtain ideological slant scores
for individual outlets as well as system-wide scores for \nn{} and \bb{}.

We use the DW-NOMINATE scores for the U.S.~Congress \cite{lewis2004measuring} as
measures of legislators' ideological locations\footnote{Based on their method,
    each member's ideological point is estimated along two dimensions.  Previous
    research has shown that -- the first dimension reveals standard left-right
    or economic cleavages, and the second dimension reflects social and
    sectional divisions.  In this paper we use only the first dimension.}.
The estimates are based on the history of roll call votes by the members of
Congress and have been widely used in political science studies and related
fields.  We classify each legislator as either ideologically-left or -right,
based on the sign of their estimates\footnote{Estimates for the 111th Congress
    are available at: \url{http://voteview.spia.uga.edu/dwnomin.htm}}. We then
calculate the ideological slant score $\theta_{ik}$, $k\in\{\mbox{Left,
        Right}\}$ for each outlet $i$
with $k=\mathrm{Left}$ so that $\theta_{i}>0$ indicates outlet $i$ is more
likely to be Left-slanted.

Our ideological slant measurements are also summarized in Table
\ref{slant-scores}. We find this measure is highly correlated with the party
slant measurement (with Pearson correlation $r=0.958$ and $p<10^{-5}$). This
suggests that, while party members may be found at different positions in the
left-right spectrum, media outlets tend to pick legislators who are
representatives of the two parties' main ideologies, such as Left-wing Democrats
or Right-wing Republicans.

\subsubsection{Gender}
Gender is also treated as a dichotomous variable, 
where $\theta_{i}>0$ indicates that the coverage of outlet $i$ favors male
legislators. The results, summarized in Table \ref{slant-scores}, show that
blogs have a slightly stronger female-slant than news. However, when considering
the population baseline, the slant for both media is significant for the Senate
but nearly insignificant for the House. The gender composition in both chambers
is similar -- 20\% of the members are women. The differences in the estimates
based on different baselines reflect a very different voter population
represented by the female/male legislators in both chambers. 

\subsubsection{Region}
We consider region as a categorical variable. For each legislator, the state or
territory of his or her district is used. The region slant is calculated like
the front-runner slant: the slant score $\theta_i$ is defined as per
Eqs.~\ref{glor} and \ref{glor-var}, where $k\in\{1,2,...,S\}$ is the rank index
for one of the $S$ states in the US, ordered by the number of references
received from outlet $i$. The results are again summarized in Table
\ref{slant-scores}. Overall, news outlets show a much stronger regional bias
than blogs. The negative slant scores in the House, based on the population
baseline, indicate outlets' favor those representatives from more populous
states.

\section{Examining Coverage}
As mentioned earlier, the slant scores of media outlets are calculated based
only on the quantity of references to legislators, and are independent of the
coverage content. In this section, we examine two intrinsic aspects of this
coverage, the hyperlinks between outlets and the sentiments of the textual
content, as related to the party slants.

\subsection{Links}
We extract the hyperlinks embedded in each news article or blog post and study
how media outlets with different slants link to one another. Using the sign of
the party slant score $\theta_p$, we divide \nn{} and \bb{} into four sectors:
D-slanted news, R-slanted news, D-slanted blogs, and R-slanted blogs.

Table \ref{tbl:hyperlinks} shows the prevalence of links among the four sectors.
Each entry $(i,j)$ represents the total number of hyperlinks from outlets in
category $i$ pointing to the articles of outlets in category $j$. The linking
pattern exhibits interesting phenomena: first and the most obvious
characteristic between the two media is that news outlets have far fewer
hyperlinks in their articles compared with blog posts.  Blogs with more
hyperlinks can also be seen as second-hand reporters or commentators in response
to some news articles and other blog posts. Second, articles in the D-slanted
outlets, including news and blogs, are more likely to be cited, including by
outlets with the opposite slant. For example, the R-slanted blogs have a large
number of hyperlinks to the D-slanted news outlets. Third, the matrix shows a
strong assortativity~\cite{newman2003mixing} in the D-slanted community -- the
D-slanted blogs are more likely to cite articles from D-slanted news and blogs
than the R-slanted blogs are to cite R-slanted news and blogs. In fact, linking
patterns among the R-slanted community appear to be disassortative.   It would
be interesting to compare our results with those of Adamic, \emph{et
    al.}~\shortcite{adamic2005political}.
\begin{table}[t!]
    \centering 
    \caption{\label{tbl:hyperlinks}The strength of hyperlinks among \nn{} and
        \bb{} with Democrat or Republican slants. Each entry $(i,j)$ represents
        the total number of hyperlinks from category $i$ to $j$.}\footnotesize
    \begin{tabular}{rcccc}
    \toprule              
	          & News (R) & News (D) & Blogs (R) & Blogs (D) \\
	\midrule
	News (R)  & 99       & 125      & 68        & 67        \\
	News (D)  & 84       & 234      & 69        & 152       \\
	Blogs (R) & 256      & 500      & 287       & 293       \\
	Blogs (D) & 298      & 895      & 299       & 623       \\
    \bottomrule
    \end{tabular}
\end{table}

\subsection{Texts}
Our slant estimation is based on how many times an outlet references a
legislator, regardless of positive or negative attitude. Without any sentiment
information, the estimated scores need to be interpreted carefully: a
significant slant score only reflects the existence of bias, but not the
polarity (if any) of such bias. This subsection describes our attempt to study
sentiment information within the media. We employ the OpenAmplify
APIs\footnote{\url{http://community.openamplify.com/}} to extract the sentiment
information of each reference. The APIs return, for each article, the detected
name entities and the sentiment values associated with the entities. We derive
sentiment information for (outlet, legislator) pairs by matching legislator
names to the names detected in each article, then aggregate the sentiment scores
associated with these legislators over all of the outlet's articles.  The
sentiment scores for parties can be derived from the scores received by party
members.

\begin{figure}
  \centering
  {\includegraphics[trim=0 8 0 0]{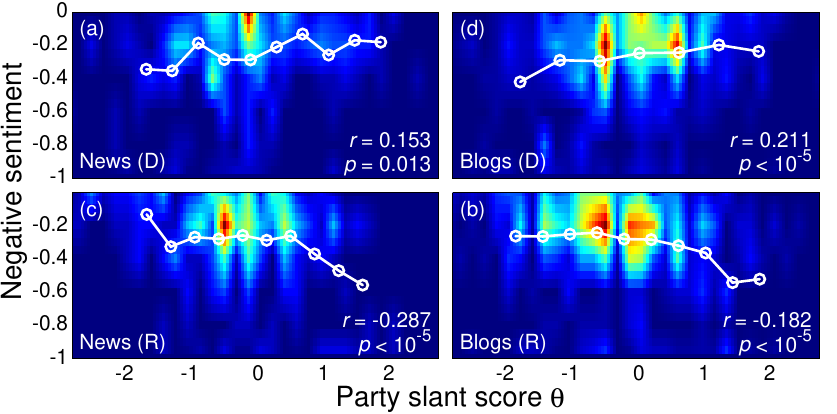}}
  \caption{\label{fig:senti} Joint probability density for negative sentiment
      and party slant score.  Solid line is the averaged trend. We see that
      D-slanted media are positively correlated with $\theta$ while R-slanted
      media are negatively correlated ($r$: correlation coefficient; $p$:
      $p$-value).}
\end{figure}

Figure \ref{fig:senti} shows the probability density of the resultant negative
sentiment scores against the party slant scores. 
The results show a weak correlation between sentiment values and the party slant
scores. Outlets' sentiments for Democratic legislators are positively correlated
to their slant scores, while sentiments for Republican legislators are
negatively correlated. This suggests the outlets with slants to a particular
party tend to mention that party less negatively. Then tendency is easier to
discover in \bb{} than in \nn{}, but this can be caused by differences in the
use of language rather than the level of bias.

\section{Modeling the reference-generating process}

What are the underlying mechanisms governing how \nn{} and \bb{} choose to
reference legislators? Are there similarities or differences between these two
media?  We propose to use a simple generative model~\cite{bagrow2008phase} for
the probability $P(n)$ that a legislator is referenced a total of $n$ times.
Comparing the results of the model's isolated mechanism with the actual data
will give intuition about factors contributing to the observed $P(n)$.

The model is as follows. Initially ($t=0$), we assume\footnote{This initial
    condition differs from the flat start of Bagrow, et
    al.~\shortcite{bagrow2008phase}, with important consequences for finite-time
    models.} a single reference to some legislator $k'$ such that  $n_k(0) =
\delta(k,k')$, for all $k$. At each time step the media (\nn{} or \bb{}) selects
a random legislator to reference in an article.  With probability $q$, however,
the media rejects that legislator and instead references a legislator with
probability proportional to his or her current coverage.  That is, at each time
step $t$, $n_k(t+1) = n_k(t) +1$ occurs with probability $p_k(t)$:
\begin{equation}
p_{k}(t)=\begin{cases}
1 / \left|\Vl \right| & \text{with prob.~$1-q$ }, \\
n_k(t) / \sum_{k'}n_{k'}(t) & \text{with prob.~$q$}. 
\end{cases}
\end{equation}
This captures the intuitive ``rich-get-richer'' notion of fame, while the
parameter $q$ tunes its relative strength. Those legislators lucky (or
newsworthy) enough to be referenced early on are likely to become heavily
referenced, since they have more opportunities to receive references, especially
as $q$ increases. Since one reference is handed out at each timestep, the total
number of references measured empirically fixes the timespan over which the
model is run; $\left|\Vl\right|$ is also fixed, so the model has one parameter,
$q$. Asymptotically ($\left|\Vl\right|\to\infty$), this model gives a pure power
law $P(n) \sim n^{-1-1/q}$ for all $q>0$~\cite{bagrow2008phase}. The
distribution of $n$ is more complex for finite $\left|\Vl\right|$, however,
obtaining a gaussian-like form for $q<1/2$ and a heavy-tailed distribution for
$q>1/2$.

\begin{figure}
  \centering
  {\includegraphics[trim=0 10 0 5]{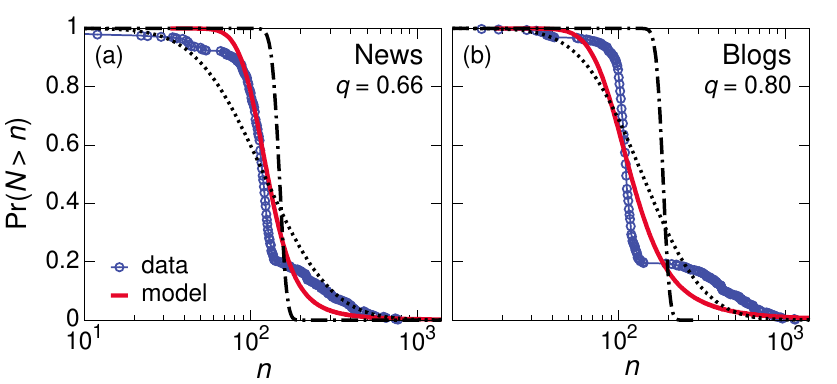}}
  \caption{\label{fig:model} The generative model for the distribution of
      references $n$ per legislator.  The larger value of $q$ for \bb{}
      indicates that they are more driven by the rich-get-richer mechanism than
      \nn{}, although both distributions are heavy-tailed. Dashed lines
      indicated fitted poisson and log-normal distributions, for comparison.}
\end{figure}

Figure~\ref{fig:model} compares the observed $P(n)$ with that generated using
the model process.  We observe good qualitative agreement, better than fitted
poisson or log-normal distributions, although there is a slight tendency to
overestimate popular legislators and underestimate unpopular legislators.  The
empirical distributions also exhibit a slight bimodality, perhaps due to the
2010 election, that is not captured by the model. The larger value of $q$ for
\bb{} than for \nn{} provides evidence that \bb{} collectively are more driven
by a rich-get-richer selection process than \nn{}, although this may not hold at
the individual outlet level.

The measures of front-runner slant indicate that \nn{} have a stronger
front-runner bias than \bb{}. This seems to conflict with the reference
generating model, which showed that blog behavior is more explainable by the
rich-get-richer mechanism ($q$ is larger for \bb{} than for \nn{}). However, we
argue that the measures and the model are in fact consistent, since the model
only treats the aggregate of the entire media class -- the stronger front-runner
bias in \nn{} outlets means that each outlet is more likely to reference their
own \textit{intrinsic} set of front-runners, which may be different from
others'; for \bb{}, the ``stickiness'' of their individual set of front-runners
is weaker and hence over time globally popular front-runners are more likely to
emerge. Further examination of this argument would be to explicitly model the
bias of individual outlets.

This one-parameter model neglects a number of dynamical features that may be
worth future pursuit.  For example, generalizations may be able to explain
temporal dynamics of the references, the joint distributions $n_{ik}$ between
media outlet $i$ and legislator $k$, etc.

\section{Discussion and Open Issues}
Our results show that \nn{} and \bb{}, in aggregate, have only slightly
different slants in terms of party and ideology. However, the dynamics of the
party slant measures suggest blogs are more sensitive to exogenous shocks, such
as the mid-term election. Our observations were made over a short, four-month
timeframe, yet long-term, continuous tracking of slant dynamics would be
necessary to reveal any consistently different dynamical behavior between the
two media.

Our measures and model are solely based on the quantity of coverage. We have
conducted preliminary sentiment analysis using an off-the-shelf tool and
compared the extracted sentiment results with our measures. The results suggest
a weak connection between the quantity and semantics of referencing a subject.
It would be worth investigating the accuracy of sentiment detection on different
media content and how sentiment analysis can be used to identify bias from
texts. In addition, critical content analysis (which examines not only the text
but also the relationship with audience) and multivariate analysis (since
multiple types of slants are inter-related) may be leveraged for further
analysis.

\section{Conclusion}
In this paper, we develop system-wide bias measures to quantify bias in
mainstream and social media, based on the number of times media outlets
reference to the members of the 111th US Congress. In addition to empirical
measurements, we also present a generative model to explore how each media's
global distribution of the number of references per legislator evolves over
time. We observe that social media are indeed more social, i.e.~more affected by
network and exogenous factors, resulting in a more heavily-skewed and uneven
distribution of popularity.  Perhaps, there are more voices than ever, but many
are echoes.

We plan to continue work along the lines discussed in the previous section, such
as long-term tracking of slant dynamics in the two media, modeling individual
outlets' biases, and leveraging content analysis and multivariate analysis.

\subsection*{Acknowledgments}
We thank F.~Simini and J.~Menche for many useful discussions, and gratefully
acknowledge support from NSF grant \# 0429452. Any opinions, findings, and
conclusions or recommendations expressed in this material are those of the
authors and do not necessarily reﬂect the views of the NSF.

\end{document}